\documentclass[twocolumn,showkeys,preprintnumbers,amssymb]{revtex4}

\usepackage{graphicx}
\usepackage{dcolumn}
\usepackage{bm}

\usepackage{fancyhdr}

\usepackage{pslatex}

\usepackage{t1enc}
\usepackage{psfrag}
\newcommand{\rhoc}{\rho_{\scriptscriptstyle\rm c}}

\begin{document}

\title{\Large Heterogeneities in Aging Models of Granular Compaction} 
\author{Jeferson J. Arenzon}
\email{arenzon@if.ufrgs.br}
\affiliation{Instituto de F{\'\i}sica, Universidade 
         Federal do Rio Grande do Sul\\ 
         CP 15051, 91501-970 Porto Alegre RS, Brazil}

\received{}
\begin{abstract}
Kinetically constrained models (KCM) are systems with trivial thermodynamics 
but often complex dynamical behavior due to constraints on the accessible 
paths followed by the system. Exploring these properties, the Kob-Andersen (KA) 
model was introduced to study the slow dynamics of glass forming liquids and 
later extended to granular materials. In this last context, we present new
results on the heterogeneous character of both in and out of equilibrium
dynamics, further stretching the granular-glass analogy.


\end{abstract}
\maketitle

\thispagestyle{fancy}
\setcounter{page}{0}

\section{Introduction}

In recent years, the analogy between structural glasses and dense granular systems 
has become deeper and extensively 
explored~\cite{KnFaLaJaNa95,NoKnPoJaNa97,LiNa98,NoKnBeJaNa98,Kurchan99,JoTkMuJa00,NiDuPo00,BaKuLoSe00,HeLaLiNa01,DaGr01b,DaGr01c,BeCuIg01,PhBi02,SiErGrHaLe02,MaKu02,PoBeNi03,KaDe04,MaDa05,MaBrEd05,DaMaBi05,SoWaMa05,GoSw05}. 
As temperature is lowered, or density increased, respectively,
both systems undergo a glass, or jamming, transition, where the
relaxation times dramatically increase.
Yet, in spite of all evidence of phenomenological similarity, the two systems are fundamentally 
different, for example, in the length scale of its components (molecules versus
macroscopic particles) and the role of thermal energy (none, in the case of granulars). 
Because the thermal energy is too small to induce movement in these 
macroscopic particles, energy should be externally supplied, for example, by
vibrating, tapping, shearing, rotating, etc, the system, 
in order for a granular system explore its configurational space.

Recently, the role of dynamical heterogeneities in the complex dynamics of glassy
systems has been addressed~\cite{Sillescu99,Ediger00,Richert02,Andersen05}. The 
glass transition seems to be purely dynamical, with
no increasing static correlation length as the system approaches the 
transition, differently from the usual critical slowing down. 
Thus, the increase in relaxation times seems to be related to a
diverging dynamic correlation 
length~\cite{YaOn98,BeDoBaGl99,GaCh02,BeGa03b,Berthier04b,BiBo04,ToWyBeBiBo05}, 
associated to 
the increasing number of particles whose displacements become dynamically
correlated, the heterogeneities, that develop during the evolution of the system. 
In Ref.~\cite{ArLeSe03} we showed, at the level of a simple, kinetically
constrained model~\cite{RiSo03}, 
that dynamical heterogeneities seem to play the same role in granular systems
as they do in structural glasses, with an increasing length scale as the system
approaches the jamming transition. This has also been observed in
other recent works~\cite{DrHaReRe05,LeBeSt05,SiLiNa05,DaMaBi05}. However,
the precise role played by these structures and the associated lengths, on the
dynamics of granular and colloidal systems, is yet to be understood.

The jamming transition has also been studied in connection with the 
concept of dynamically available volume (DAV)~\cite{MeBa00,LaReMcGrTaDa02}.
Apart from having enough nearby empty space~\cite{BoGe97}, a 
particle is mobile if, for the particular type of model considered here, 
the displacement is allowed by the kinetic constraints. Empty sites that are able to receive a 
neighboring mobile particle are called {\it holes}.  Holes can be 
classified as either connected or not, the former being those that, by allowing 
a particle to jump into it may eventually facilitate the movements of all 
particles in the system. On the other hand, nonconnected holes leave a backbone
of blocked particles.  Close to dynamical arrest, the density of
connected holes decreases with the density of particles and is related with
the inverse of the bootstrap length~\cite{GrLaBrDa04,GrLaBrDa05}, the average
distance between two connected holes, in
bootstrap percolation. In turn, this length is 
associated with the transport properties of the system~\cite{JaKr94,ToBiFi04,ToBiFi05a}.

Here we make an initial attempt of studying how these holes behave
when the system is externally driven and falls out of equilibrium, in particular, their role
during the compaction regime of granular systems, what is connected with the
voids distribution measured experimentally~\cite{RiPhBa03}. Although we do not distinguish
at this stage connected from disconnected holes, this would be
important in order to fully understand the microscopic compaction
mechanism.

\section{Kob-Andersen model}

The Kob-Andersen model~\cite{KoAn93} is one of the simplest models
describing the complex dynamics of glassy and granular systems.
It consists of a lattice gas of $N$ particles, each site being either empty
or occupied by one particle, with no static interactions between them,
i.e., ${\cal H}=0$. In addition, a kinetic constraint~\cite{RiSo03}
should be satisfied in order to allow the displacement of a particle
to an empty neighboring site: there should be fewer than $m$  
occupied nearest neighbors {\it before} and {\it after} the move. This kinetic 
rule is time-reversible and detailed balance is satisfied.
If the constraint is obeyed, the particle is said to be mobile and the
companion vacant site is said to be a hole. At high densities, the 
dynamics slows down because the reduced free-volume makes it harder for 
a particle to satisfy the dynamic constraints. These constraints were
introduced to mimic the cage effect where, due to geometrical
effects, the displacements of the particles are hindered
by their neighbor particles.

Although on hypercubic lattices this model presents a jamming transition 
only at full occupancy~\cite{ToBiFi04}, with a super Arrhenius behavior of the
relaxation time, for finite lattices the bootstrap
length may become larger than the system size $L$. Thus, for a finite
sample, the system may be frozen due to the lack of connected holes,
implying the existence of a dynamical critical density $\rhoc(L)<1$, 
slowly increasing with $L$. At this point, it is observed that the diffusivity 
falls to zero as a power law~\cite{KoAn93}, $D(\rho)\sim (\rhoc-\rho)^{\phi}$, 
with $\rhoc$ depending, in addition, both on the lattice geometry and on the 
kinetic constraint $m$~\cite{KoAn93,ImPe00,ArLeSe03}.
Assuming such a power law form for the diffusivity, 
finite systems, both with and without gravity, are well described, 
qualitatively and quantitatively, by
a nonlinear diffusion equation~\cite{PeSe98,LeArSe01,ArLeSe03}.

Since much of dynamical properties of both structural glasses and
dense granular materials are dictated by steric constraints, we have
generalized~\cite{SeAr00} the Kob-Andersen model by including a gravitational
field.  The Hamiltonian now has a one body term, 
$\beta {\cal H} = \gamma \sum_i z_i n_i$,
where $n_i=0,1$ is the occupation variable of the $i$-th site whose
height is $z_i$, $\gamma= m g/k_{\scriptscriptstyle\rm B} T$ is the
inverse gravitational
length and $g$ is the constant gravitational field acting in the 
downward direction. We follow a continuous vibration dynamics, assuming
that the random diffusive motion of particles, produced by the
mechanical vibrations of the box, can be modeled as a thermal bath of
temperature $T$. The particles satisfying the
kinetic constraints may always move downwards while upward movements
are accepted with a probability $x =\exp(-\gamma)$, related to the 
vibration amplitude.  Particles are
confined in a closed box of bcc structure, with periodic boundary
conditions in the horizontal direction.  We set the
constraint threshold at $m=5$.  As the Markov process generated by the
kinetic rules is irreducible on the full configuration
space~\cite{SeAr00}, the static properties of the model are those of a
lattice-gas of non-interacting particles in a gravitational field, and
these can be easily computed.

\section{Heterogeneities}

\begin{figure}[floatfix]
\begin{center}
\includegraphics[width=6cm,angle=270]{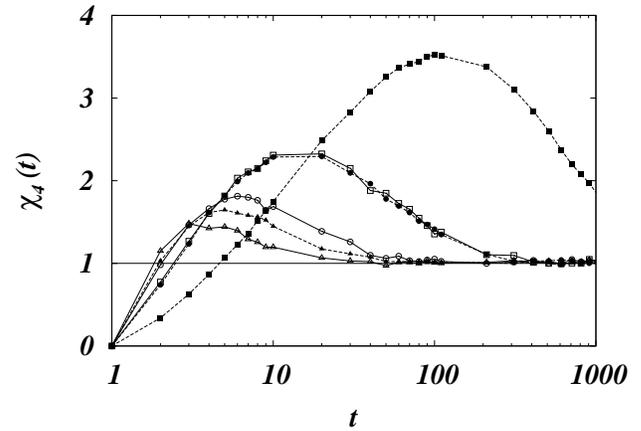}
\end{center}
\caption{ Dynamical response, eq.~\ref{eq.het}, as a function of time
         (in MCS) for different vibrations: x=0.92 (filled symbols)
         and 0.94 (empty symbols).  Different symbols stand for
         different heights: $z=5, 10$ and 15 (square, circle,
         triangle, respectively).
         The line is the asymptotic $\chi_4=1$ behavior~\cite{RiSo03}.  Notice in
         the figure the presence of two very close curves: they
         correspond to different vibrations and heights, but their
         density is the same to within numerical accuracy.  
         From \cite{ArLeSe03}.}
\label{fig.het}
\end{figure}

\begin{figure}[floatfix]

\begin{center}
\psfrag{x}{$(1-\rho)^{-1}$}
\psfrag{y}{\Large $\ln\ln\tau$}
\includegraphics[width=8.4cm]{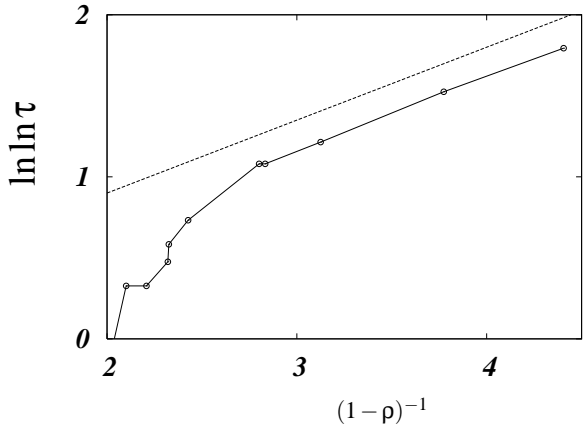}
\end{center}

\caption{Position of the peak in $\chi_4$. The growth is compatible
         with the known relaxation time~\cite{ToBiFi04}, Eq.~\ref{eq.tau}
         (dashed line).}
\label{fig.het2}
\end{figure}

Several measures for quantifying spatial heterogeneities have been
introduced for kinetic models~\cite{RiSo03}. In particular, this issue
was recently investigated in the KA model~\cite{FrMuPa02,ToBiFi05a} without
gravity using fourth-order correlation functions~\cite{GrLaBrDa04,GrLaBrDa05}. 
In Ref.~\cite{ArLeSe03} these were extended to include the non-zero gravity case.
In Fig.~\ref{fig.het} we plot the dynamical nonlinear response
\begin{equation}
\chi_4(z,t)=N \left( \left\langle q^2(z,t) \right\rangle -
\left\langle q(z,t) \right\rangle^2 \right)
\label{eq.het}
\end{equation}
where $N$ is the number of sites in the computation,
$q(z,t)=C(z,t)/C(z,0)$ and
\begin{equation}
C(z,t) =\frac{1}{N}\sum_i n_i(t)n_i(0) - \rho(z,t)\rho(z,0).
\end{equation}
For all vibrations considered, the system is in the fluid phase and
is able to achieve the asymptotic state very fast.
Notice that, differently from previous works, here we measure these 
quantities separately around each 
layer of the system (to improve the averages, $i$ runs over all sites 
in the $z$, $z-1$ and $z+1$ layers). In other words, we are probing
horizontal, microscopic heterogeneities, not the macroscopic ones
due to the intrinsicly inhomogeneous vertical profile. 
Analogously to what happens in the KA model without
gravity and in other glassy systems, the peak is shifted to higher
times and gets larger as the density increases (the lower is $z$, the
greater is the density), which is an indication of cooperative
dynamics, as larger clusters have more difficulty to
respond to a perturbation. As these measures are
done for vibrations above the apparent jamming threshold, the
behavior of the system should not be affected by the finite size
shortcomings discussed above and one is able to obtain
the true, infinite size behavior. Indeed, the growth of the 
peak is compatible with the known relaxation time~\cite{ToBiFi04},
\begin{equation}
\tau \sim \exp \exp \left( \frac{c}{1-\rho} \right)
\label{eq.tau}
\end{equation}
as can be seen in Fig.~\ref{fig.het2}. Analogous results can be
obtained also from the peak of the Kovacs hump~\cite{ArSe04} and
persistence times distribution. 
Interestingly, $\chi_4$ only depends on $z$ through its local density,
$\chi_4(z,t)=\chi_4(\rho(z),t)$: 
Fig.~\ref{fig.het} shows that two curves corresponding
to different heights and vibrations ($z$ and $x$), but having almost the same density
(within numerical precision), coincide.

\begin{figure}[floatfix]
\begin{center}
\includegraphics[width=6cm,angle=270]{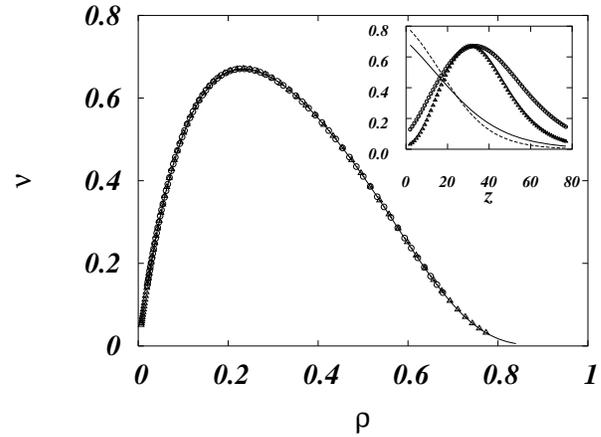}
\end{center}
\caption{Density of holes as a function of density 
for $x=0.92$ (triangle) and 0.94 (circle), along with the
results for the KA model without gravity (line) in a bcc lattice
of height $4L$, with $L=20$. Inset: holes
profile (symbols) along with the 
corresponding density profiles (lines). The higher $x$ corresponds
to the flatter density profile and to the broader holes profile.}
\label{fig.holes}
\end{figure}

A more direct measure of inhomogeneity comes from the mobile
particles or, equivalently, the holes at different heights,
without distinguishing, at this stage, between connected and
non-connected holes.
In the inset of Fig.~\ref{fig.holes}, we show the density and holes 
profiles, $\rho(z)$  and $\nu(z)$, respectively,
still in the fluid phase, as a function of $z$ for two different 
vibrations $x$. The density decreases with height and the stronger the 
vibration, the flatter is the
density profile and the broader is $\nu(z)$, that is, the more vertically
homogeneous the system is. Indeed, for $x=1$ (no preferential
direction) both profiles are flat.
 Notice also that as $x$ decreases, the holes 
concentrate in the region that will form the interface once 
the system goes out of equilibrium. 
The interesting result that can be seen in Fig.~\ref{fig.holes} appears 
when we eliminate $z$ and plot $\nu(z)$ as a function of $\rho(z)$: 
the data for different values of $x$ perfectly collapse onto a single 
master curve. Moreover, they are on top of the points for the KA model 
without gravity. In other words: even 
if the system is no longer homogeneous, the local density of holes in 
equilibrium only depends on the local density, and not on the whole 
profile, or on the vibration or the height. 
This, together with the analogous result for $\chi_4$, 
if general, is an
important property: as some of the quantities depend only on the local 
density, not on the whole profile, there is further support for even simpler,
one dimensional models as well as local density approximations~\cite{PeSe98,LeArSe01,ArLeSe03}. 
>From the point of view of simulation, it is a
fast way to obtain, in just one simulation, the whole profile for
systems without gravity.

\begin{figure}[floatfix]
\begin{center}
\includegraphics[width=6cm,angle=270]{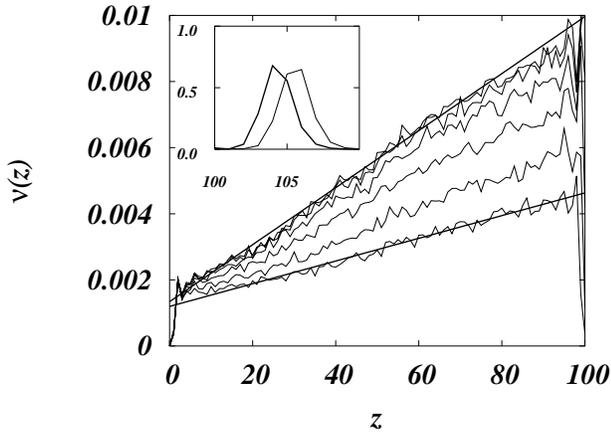}
\end{center}
\caption{Hole profile at different times (smaller time at the top). 
The profile is made of two clearly separated
parts: $i$) the interface (inset) contains almost all holes in the
system and moves to the left as the system becomes compact; $ii$) the
bulk has a much smaller hole fraction, that increases linearly with
height. Separating these regions we have a few layers near $z=100$
that are almost hole-free.
The declivity slowly decreases with time. Notice also the difference
on the vertical scales.}
\label{fig.holespro}
\end{figure}

\begin{figure}[floatfix]
\begin{center}
\includegraphics[width=6cm,angle=270]{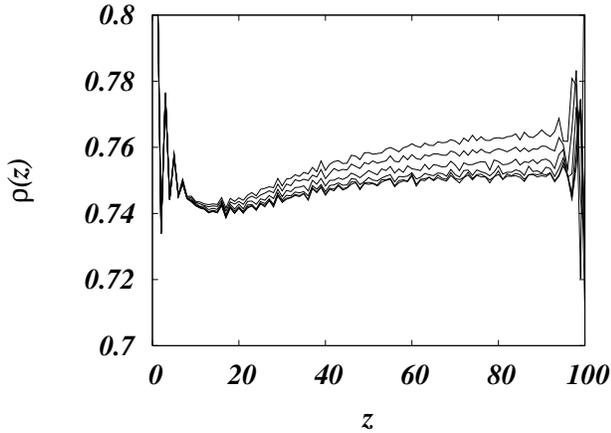}
\end{center}
\caption{Bulk density profile at different times (smaller time at the bottom). 
The region that compactifies the most is the denser one, close to the top, due to
a larger number of holes.}
\label{fig.rhopro}
\end{figure}

Upon decreasing the vibration, a finite system enters in the aging
regime and the density profile develops two very distinct 
regions~\cite{LeArSe01}: an almost flat plateau at $\rhoc$ for $z<z_0$ 
and a density decreasing region (interface) for $z>z_0$. The
position of this plateau is size dependent and slowly increases
with $L$, similar to $\rho_c$. 
Analogously, the hole profile also has two corresponding
regions (see Fig.~\ref{fig.holespro} and the inset): at the interface 
where most holes are localized and in the bulk, where their number
is much smaller. These regions are separated by a dip with a few layers 
width, where there are almost no holes, corresponding to the dense layer
seen in the density profile near the interface, at the top of the
granular pile
(this region also seems to become more localized with time), as
seen in Fig.~\ref{fig.rhopro}. 
At the interface, the profile is strongly peaked, and moves to the
left, accompanying the interface, as the system ages (compactifies). 
On the other hand, in the bulk, the profile 
is linear, with a tangent that slowly decreases in time (although also 
compatible with the inverse logarithm law, a good fit is obtained with 
$t^{-0.25}$). This evaporation of holes is the direct mechanism of
the compactification process. Also, because of the almost hole-free 
layers between the bulk and the interface, it is very hard to exchange 
particles between the bulk and the interface. The interesting result 
here is that this process is
inhomogeneous: it is faster at the topmost region of the
bulk (just below the dense layer) and slower at the bottom of the 
system~\cite{Sellitto02a}, as seen in Fig.~\ref{fig.rhopro}. The oddity comes from the fact that
the compaction is faster where the density is higher, the region
where one would expect a 
slower evolution and a smaller number of holes. Instead, the evolution
is faster and the number of holes is higher. Finally, it should be
remarked that the small positive gradient seen in the density
profile is consistent with
experimental results~\cite{KnFaLaJaNa95,PhBi02}, although its
sign is model dependent~\cite{PhBi01}.

\begin{figure}[floatfix]
\begin{center}
\includegraphics[width=6cm,angle=270]{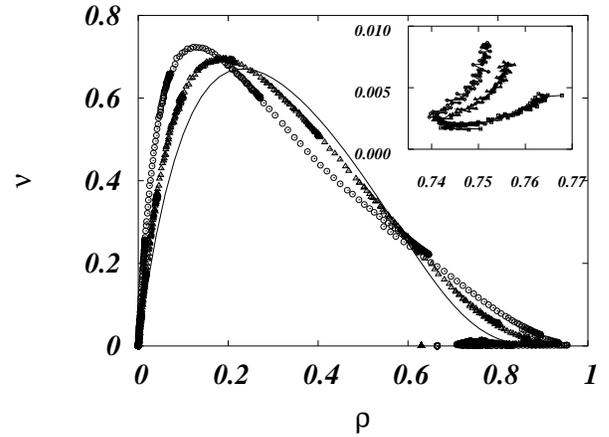}
\end{center}
\caption{Density of holes as a function of density for several
values of $x$ in the aging regime (0.2, circle, and 0.4, triangle), 
along with the equilibrium results for the KA model (line).
Differently from the fluid phase,
here there is no collapse of the curves onto the equilibrium
curve. Notice that there are some points with $\rho>\rho_c$ that
come from the oscillatory/dense layer regions.
It is interesting to notice that in the case without gravity, the number
of holes in the high-$\rho$ region is below the equilibrium one, while
here it is above. Inset: Holes coming from the bulk for $x=0.2$ and 
three different
times (from left to right: $t=1072$, 10974 and 99999) as a
function of density. Differently from what would be
expected, this an increasing function of density. In these plots we
do not take into account the oscillating layers near the bottom and
near the dense layer, only the bulk layers ($6<z<95$).}
\label{fig.holesaging}
\end{figure}

In Fig.~\ref{fig.holesaging} the parametric plot of $\nu$ versus 
$\rho$ in the out-of-equilibrium, aging regime is shown for several
vibrations $x$ along with the equilibrium curve. Notice
that despite the fact of not being stationary, the points,
for several times (but the same $x$), roughly fall on a master curve,
while different vibrations no longer collapse onto the same curve.
However, these $x$-dependent universal curves 
no longer correspond to the equilibrium one, and as the
vibration increases, there is a drift toward the equilibrium curve.
The two regions seen in the $\nu$ profile, Fig.~\ref{fig.holespro},
contribute differently for the curve in Fig.~\ref{fig.holesaging}: 
the bulk forms the high $\rho$, small $\nu$ region
near the horizontal axis (enlarged in the inset), while the interface 
generates the rest of the curve.
In the inset of Fig.~\ref{fig.holesaging} we can see that the
bulk behavior is
the opposite of what would be expected at high densities: as the
density increases the corresponding holes also increase! This
explains why the density profile evolves faster near the dense
layer where the density is higher: the greater the number of
holes, the easier it is to compactify.
Moreover, for different times, the curves no longer collapse (notice
that the inset of Fig.~\ref{fig.holesaging} shows data for different
times but same vibration $x$).
Thus, differently from the holes from the interface, there
is no universal curve for the holes in the bulk. Interestingly,
besides having the opposite $\rho$ dependence, for larger times 
the curves move away from the equilibrium one
(that is one order of magnitude above).
Even the interface behavior is puzzling: layers whose density is
small would be expected to be in local equilibrium. However, the
hole distribution only coincides with the equilibrium one
for $\rho\to 0$.

\section{Conclusions}

In summary, besides briefly reviewing the first application of the
dynamical susceptibility $\chi_4$ in the context of granular
compaction~\cite{ArLeSe03}, we also extended the notion of dynamically available
volumes for this externally driven, out of equilibrium situation. 
This is an additional similarity  between structural glasses
and granular systems, while shedding some light on the microscopic 
mechanisms responsible for the slow dynamics close to these transitions. 
Interestingly, the stationary behavior
of the Kob-Andersen model with gravity is characterized by the local
density, in spite of the macroscopically inhomogeneous density profile and the 
vibration imposed: layers with the same density presents the
the same time dependence of $\chi_4(z,t)$ as well as the
same density of holes.  In the aging, compaction regime,
this is no longer the case, although there is still a
certain degree of universality in the behavior
of the total number of holes from the interface, reflected on the collapse, 
at different times (but
different densities), of all points onto a roughly universal curve
that depends, in its turn, on the vibration (temperature). However,
how connected and non-connected holes contribute to these profiles are
yet to be investigated.

In addition, there are still many issues that deserve a closer inspection.
Important information can also be obtained from the different 
sectors of the $\chi_4(t)$ function~\cite{ToWyBeBiBo05}. However,
the range of time/density considered here is still small to
resolve these different sectors. The distinction between connected and
disconnected holes should also be made, as the former should be
much more important in the compaction mechanism, along with
the study of their spatial distribution~\cite{LaGrBrSeDa05}.
Correlations may be also defined considering only the holes, with
the associated dynamical susceptibility.

\begin{acknowledgments}
This work was partially supported by the Brazilian agencies
CNPq, CAPES and FAPERGS. I acknowledge conversations with Mauro Sellitto, 
Yan Levin, Paolo di Gregorio and Aonghus Lawlor. Finally, I very much
profited of the Abdus Salam ICTP (Italy) associationship
support and kind environment during the last years, where part 
of this research was carried on. 
\end{acknowledgments}

\bibliographystyle{prsty}

\bibliography{sg}

\end{document}